\documentclass[floatfix,prd,showpacs,twocolumn,preprintnumbers,amsmath,amssymb]{revtex4}

\usepackage{graphicx}
\usepackage{dcolumn}
\usepackage{amsmath}
\usepackage{epsfig}
\usepackage{subfigure,hyperref}
\usepackage[normalem,normalbf]{ulem}

\newcommand{\be}{\begin{equation}}
\newcommand{\ee}{\end{equation}}

\DeclareRobustCommand\openone{\leavevmode\hbox{\small1\normalsize\kern-.33em1}}
\newcommand\ie{{\it i.e. }}

\def\nbC{{\mathchoice {\setbox0=\hbox{$\displaystyle\rm C$}%
\hbox{\hbox to0pt{\kern0.4\wd0\vrule height0.9\ht0\hss}\box0}}
{\setbox0=\hbox{$\textstyle\rm
C$}\hbox{\hbox to0pt{\kern0.4\wd0\vrule height0.9\ht0\hss}\box0}}
{\setbox0=\hbox{$\scriptstyle\rm
C$}\hbox{\hbox to0pt{\kern0.4\wd0\vrule height0.9\ht0\hss}\box0}}
{\setbox0=\hbox{$\scriptscriptstyle\rm C$}\hbox{\hbox to0pt{\kern0.4\wd0\vrule
height0.9\ht0\hss}\box0}}}}
\def\nbQ{{\mathchoice {\setbox0=\hbox{$\displaystyle\rm
Q$}\hbox{\raise 0.15\ht0\hbox
to0pt{\kern0.4\wd0\vrule height0.8\ht0\hss}\box0}}
{\setbox0=\hbox{$\textstyle\rm Q$}\hbox{\raise
0.15\ht0\hbox to0pt{\kern0.4\wd0\vrule height0.8\ht0\hss}\box0}}
{\setbox0=\hbox{$\scriptstyle\rm
Q$}\hbox{\raise 0.15\ht0\hbox to0pt{\kern0.4\wd0\vrule
height0.7\ht0\hss}\box0}}
{\setbox0=\hbox{$\scriptscriptstyle\rm Q$}\hbox{\raise 0.15\ht0\hbox
to0pt{\kern0.4\wd0\vrule
height0.7\ht0\hss}\box0}}}}
\def\nbT{{\mathchoice {\setbox0=\hbox{$\displaystyle\rm
T$}\hbox{\hbox to0pt{\kern0.3\wd0\vrule
height0.9\ht0\hss}\box0}} {\setbox0=\hbox{$\textstyle\rm
T$}\hbox{\hbox to0pt{\kern0.3\wd0\vrule
height0.9\ht0\hss}\box0}} {\setbox0=\hbox{$\scriptstyle\rm
T$}\hbox{\hbox to0pt{\kern0.3\wd0\vrule
height0.9\ht0\hss}\box0}} {\setbox0=\hbox{$\scriptscriptstyle\rm T$}\hbox{\hbox
to0pt{\kern0.3\wd0\vrule height0.9\ht0\hss}\box0}}}}
\def\nbS{{\mathchoice {\setbox0=\hbox{$\displaystyle     \rm
S$}\hbox{\raise0.5\ht0%
\hbox to0pt{\kern0.35\wd0\vrule height0.45\ht0\hss}\hbox
to0pt{\kern0.55\wd0\vrule
height0.5\ht0\hss}\box0}} {\setbox0=\hbox{$\textstyle        \rm
S$}\hbox{\raise0.5\ht0%
\hbox to0pt{\kern0.35\wd0\vrule height0.45\ht0\hss}\hbox
to0pt{\kern0.55\wd0\vrule
height0.5\ht0\hss}\box0}} {\setbox0=\hbox{$\scriptstyle      \rm
S$}\hbox{\raise0.5\ht0%
\hboxto0pt{\kern0.35\wd0\vrule height0.45\ht0\hss}\raise0.05\ht0%
\hbox to0pt{\kern0.5\wd0\vrule height0.45\ht0\hss}\box0}}
{\setbox0=\hbox{$\scriptscriptstyle\rm
S$}\hbox{\raise0.5\ht0%
\hboxto0pt{\kern0.4\wd0\vrule height0.45\ht0\hss}\raise0.05\ht0%
\hbox to0pt{\kern0.55\wd0\vrule height0.45\ht0\hss}\box0}}}}
\def\nbZ{{\mathchoice {\hbox{$\sf\textstyle Z\kern-0.4em Z$}}
{\hbox{$\sf\textstyle Z\kern-0.4em Z$}}
{\hbox{$\sf\scriptstyle Z\kern-0.3em Z$}}
{\hbox{$\sf\scriptscriptstyle Z\kern-0.2em Z$}}}}
\begin{document}

\title{Optimization of the derivative expansion in the nonperturbative renormalization group}

\author{L\'eonie Canet}
\email{canet@lpthe.jussieu.fr}
\author{Bertrand Delamotte}
\email{delamotte@lpthe.jussieu.fr}
\author{Dominique Mouhanna}
\email{mouhanna@lpthe.jussieu.fr}

\address{Laboratoire de Physique Th\'{e}orique et Hautes
Energies, CNRS UMR 7589, \\  Universit\'{e}
Pierre et Marie Curie Paris 6, Universit\'{e} Denis Diderot Paris 7, 2 place Jussieu, 75252 Paris Cedex 05
France}
\author{Julien Vidal}
\email{vidal@gps.jussieu.fr}
\address{Groupe de
Physique des Solides, CNRS UMR 7588,\\ Universit\'{e} Pierre et Marie Curie Paris 6, Universit\'{e} Denis Diderot Paris 7, 2 
place Jussieu, 75251
Paris Cedex 05 France}

\begin{abstract}
 We study the optimization of nonperturbative renormalization group
 equations truncated both in fields and derivatives. On the example of
 the Ising model in three dimensions, we show that the Principle of
 Minimal Sensitivity can be unambiguously implemented  at order
 $\partial^2$ of  the derivative expansion.  This  approach  allows us
 to select optimized cutoff functions and to improve the  accuracy of
 the critical exponents $\nu$ and $\eta$. The convergence of the field
 expansion is also analyzed. We show in particular  that its
 optimization  does {\it not} coincide with optimization of  the
 accuracy of the critical exponents.
\end{abstract}


\maketitle

\section{Introduction}

During     the     last      ten     years     the     Wilson-Kadanoff
 approach \cite{kadanoff66,wilson74} to the Renormalization Group (RG),
 based on the block spin concept, has been the subject of a revival in
 both Statistical Physics and  Field Theory. This originates in recent
 developments \cite{wetterich93c,ellwanger94c,morris94a} which have now
 turned  it into an efficient tool, the so-called   effective      average      action
 method \cite{wetterich93c},     allowing     to
 investigate  nonperturbative phenomena.  This
 method implements on the effective  action $\Gamma$ -- the Gibbs free
 energy -- the idea of integration of high-energy modes that underlies
 any RG approach.  The whole  method consists in building an effective
 free energy  $\Gamma_k$ at scale  $k$ for the high-energy  modes that
 have been integrated out and in  following its evolution with the scale $k$
 through an exact equation \cite{wetterich93c}.  The main
 drawback of this equation is that it
 cannot  be  handled  in   actual  calculations  without  truncations  of
 $\Gamma_k$.   It is  thus of  utmost importance  to know  whether the
 truncations  used provide  converged  and accurate  results. As it is
 well known,  the problem of  convergence is  also crucial  in perturbation
 theory where it requires resummation of series. Let us emphasize that this
 problem is far from being  solved in general since Borel summability,
 which  is  the  key  point   to  resum  perturbative  series,  is  not
 generically  proven  and   may  even  turn  out  not   to  hold  (see
  \cite{pelissetto01c,holovatch02} for a review). It is then  important to
 dispose of  an alternative method, not  relying on an  expansion in a
 coupling constant and thus  not requiring {\it a priori} resummation.
 Good  indications  on the  convergence  properties  of the  effective
 average action  method have been  already provided by its  ability to
 tackle   with  highly  nontrivial   problems,  such   as  low-energy
 QCD \cite{jungnickel99},   the  abelian   Higgs  model   relevant  for
 superconductivity \cite{bergerhoff96},   the   phase   diagram   of
 He$_3$ \cite{kindermann01}, frustrated  magnets \cite{tissier00,tissier00b,tissier01,tissier02b},           the
 Gross-Neveu  model  in three  dimensions \cite{rosa01,hofling02},  the
 randomly dilute  Ising   model \cite{tissier01b},   the   Kosterlitz-Thouless
 transition        \cite{grater95,gersdorff01},        etc        (see
 \cite{berges02} for a review and \cite{bagnuls01} for an exhaustive
 bibliography). 
 A systematic investigation  of the convergence and accuracy issues is
 however
 still lacking.

 We  propose  here,  on  the  example  of the three-dimensional Ising
 model,  to study the  convergence and optimization of the accuracy
of the  effective average action method truncated  both in derivatives,
at order  $\partial^2$, and  in fields.  We  study, in  particular, the
role  of the  cutoff function  used to  separate the  low-  and high-
energy modes, on the determination of the critical exponents $\nu$ and
$\eta$.

 In  section  II,  we  briefly  introduce  the basic  ideas  underlying  the
 effective average action  method. We then discuss in  section III the
 truncations  necessary to  deal with  concrete calculations.  We
  motivate, in  section IV, the use of the Principle of
 Minimal Sensitivity (PMS)  to optimize the  results.  Then, we  apply this
 technique  successively  within  the  Local  Potential  Approximation
 (LPA),  section  V,  and  at  order $\partial^2$  of  the  derivative
 expansion, section VI.

\section{The effective average action method}

Historically,  the block spin  concept was  first implemented,  in the
continuum,  on   the  Hamiltonian.  This  procedure  consists  in
separating, within the partition function, the microscopic fields into
a high- and  a low-energy part and in  integrating out the high-energy
part  to get  an effective  Hamiltonian for  the  remaining low-energy
modes.  The iteration of this  procedure generates a sequence, a flow,
of scale-dependent Hamiltonians,  parametrized by a running scale $k$, 
and describing the same long distance physics. The critical properties
are then  determined by  the behavior of  the system around  the fixed
point  of the  flow of  Hamiltonians \cite{wilson74}.  However,  due to
technical
difficulties \cite{morris94a,morris94b,morris98b,bagnuls01} this 
  nonperturbative  renormalization procedure has been mainly used as a
conceptual  basis  for  perturbative  calculations rather  than  as  a
practical  tool  to   investigate  nonperturbative  aspects  of field theories and 
critical phenomena. The  situation has changed when  it  has
been  realized,  mainly by  Ellwanger \cite{ellwanger93a,ellwanger93b,
ellwanger94a,      ellwanger94b},     Morris \cite{morris94a,morris94b,
morris94c,morris95a,     morris95b,    morris98,     morris98b}    and
Wetterich \cite{wetterich93,  wetterich93b, wetterich93c, wetterich93d,
wetterich94, wetterich96}  that, rather than the  Hamiltonian $H$, one
should consider the effective action $\Gamma$ -- the Gibbs free energy
-- as the  central quantity to be  renormalized. In the  spirit of the
original  Wilsonian formulation one builds  a  {\it  running}  effective action
$\Gamma_k$ that  only  includes high-energy  fluctuations  with  momenta
$q^2>k^2$.   This implies  that, on  the one  hand, at  the underlying
microscopic scale $k=\Lambda$, $\Gamma_k$ coincides with the classical
Hamiltonian $H$  since {\it}  no fluctuation has  yet been  taken into
account.   On the other  hand, when  the running  scale is  lowered to
$k=0$, {\it i.e.}  when {\it all} fluctuations have been integrated out, the
standard  effective  action  $\Gamma$  is  recovered.   To  summarize,
$\Gamma_k$   continuously   interpolates   between   the   microscopic
Hamiltonian $H$ and the free energy:
\begin{equation}
\left\{
\begin{array}{ll}
\Gamma_{k=\Lambda}=H\ 
\\ 
\\
\Gamma_{k=0}=\Gamma\ . 
\label{limitesgamma}
\end{array}
\right.
\end{equation}

Since,  by definition,  $\Gamma_k$  is built up from  the  high-energy
fluctuations of  the microscopic system, the low-energy  modes -- with
$q^2<k^2$ -- must be removed from the running partition function. This
is  most easily  achieved  by  adding to  the  original Hamiltonian  a
scale-dependent  mass  term  $\Delta  H_k$.  Then,  the  running
partition function with a source term writes \cite{berges02}:

\begin{equation}
{\cal Z}_k[J]=\int D\chi\ e^{\displaystyle -H[\chi]-\Delta H_k[\chi] +
J.\chi}
\label{partition2}
\end{equation}
with $J.\chi=\int d^d q\ J(q) \chi(- q)$ and
\begin{equation}
\Delta H_k[\chi]={1\over 2} \int {d^d q\over (2\pi)^d} \
{{R}}_k(q) \chi(q) \chi(- q)\ 
\label{mass}
\end{equation}
where  $\chi(q)$ is the microscopic field. In Eq.(\ref{mass}), $R_k(q)$ 
is chosen in such a way that it acts as a  cutoff function that decouples 
the low- and high- energy modes. This  imposes several constraints:
\begin{eqnarray}
&R_k(q)&\sim k^2  \hspace{0.5cm} \hbox{for} \hspace{0.5cm}  q^2\ll k^2
\label{IR}\\
\nonumber
\\
&R_k(q)&\to 0 \hspace{0.6cm} \hbox{for} \hspace{0.5cm} {q}^2\gg k^2 \ .
\label{IR2}
\end{eqnarray}  
Equation (\ref{IR})  means  that,  at  low-momentum  with  respect  to  $k$,
$R_k(q)$  essentially  acts as  a  mass, {\ie}an  infrared cutoff,  which
prevents the  propagation of the low-energy modes.   This ensures that
these  modes  do  not
contribute to  $\Gamma_k$ \footnote{Note that the  behavior of $R_k(q)$
given  in Eq.(\ref{IR})  is not  the most  general one.  For instance,
$R_k(q)$  can   be  a  power  law  cutoff,   {\it  i.e.}  $R_k(q)=q^2
(q^2/k^2)^{-a}$ in which case  the constraint becomes: $R_k(q)\gg q^2$
(for $q^2\ll  k^2$). }. Eq.(\ref{IR2}) implies that  $R_k(q)$ does not
affect the  propagation of high-energy  modes. They are  thus almost
fully  taken  into  account   in  ${\cal  Z}_k$  and, consequently, in
$\Gamma_k$.

In order to recover the limits (1), $R_k(q)$ must also satisfy:
\begin{equation}
R_k(q)\to \infty\hspace{0.5cm} \hbox{when} \hspace{0.5cm} k\to
\Lambda \hspace{0.5cm} \hbox{at fixed} \hspace{0.3cm} q
\end{equation}
which ensures that $\Gamma_{k}$ coincides with the
microscopic Hamiltonian $H$ when $k\rightarrow \Lambda$, and
\begin{equation}
R_k(q)\to 0 \hspace{0.5cm} \hbox{identically,}\  \hbox{when} \hspace{0.5cm} k\to 0
\hspace{0.5cm}
\end{equation}
which ensures that,  in the limit of vanishing $k$, one recovers
 the standard effective action $\Gamma$.
Note  that since we  are only  interested here  in the  universal long
distance  behavior and  not in quantities depending  on microscopic
details, we send $\Lambda$ to $\infty$.

The effective average action  $\Gamma_{k}$ is then defined as:
\begin{equation}
\Gamma_k[\phi]=-\ln {\cal Z}_k[J]+ J.\phi -\Delta H_k[\phi]
\label{defgamma}
\end{equation} 
where $\phi$ stands for the running order parameter $\phi_k(q)$:
\begin{equation}
\phi_k(q)=\langle \chi(q) \rangle_k={\delta \ln {\cal Z}_k[J]\over \delta
J(q)}{\bigg |}_{J=0}\ .
\end{equation}
 It follows from the definition (\ref{defgamma}) that $\Gamma_k[\phi]$
 essentially  corresponds  to the  Legendre  transform  of $\ln  {\cal
 Z}_k[J]$, up  to the mass term  $\Delta H_k$ which  allows to recover
 the limits (\ref{limitesgamma}) \cite{tetradis94}.

The effective average action $\Gamma_k$ follows an exact equation which 
controls its evolution with the running scale $k$ \cite{wetterich93c}:
\be
\partial_t \Gamma_k[\phi] = {1 \over2} \int {d^dq\over (2\pi)^d}\  {\partial_t R}_k(q)
\left\{\Gamma_k^{(2)}[\phi(q)] + R_k(q)\right\}^{-1}
\label{eqflot}
\ee
where  $t=\ln(k/\Lambda)$  and  $\Gamma_k^{(2)}[\phi]$ is  the  second
functional  derivative  of  $  \Gamma_k$  with respect  to  the  field
$\phi(q)$.  We emphasize that Eq.(\ref{eqflot})  is exact  and thus  contains all
perturbative  and nonperturbative  features of  the  underlying theory
(see \cite{tetradis94}  for technical details and \cite{berges02}
for  a review  of  the applications  of this  equation to  concrete
physical issues).

\section{Truncations of the effective average action}

Equation (\ref{eqflot})  is a functional  partial integro-differential
equation  that  has  obviously   no  known  solution  in  the  general
case.  Therefore, to  render it  tractable, one  has to  truncate the
effective action $\Gamma_k$. The  most natural truncation, well suited
to  the  study  of  the  long  distance  physics,  is  the  derivative
expansion. It consists in writing  an {\em ansatz} for $\Gamma_k$ as a
power series in $\partial \phi$. Let  us first consider the case of an
$O(N)$ invariant theory for which the {\em ansatz} at the order
$\partial^2$  writes \cite{tetradis94}:
\begin{equation}
\begin{split}
\Gamma_k[\phi]=\int d^d x\;\Bigg\{U_k\left(\rho \right) +{1\over 2} 
Z_k(\rho)\left(\partial_{\mu}\vec\phi\right)^2 +\\{1\over 4} 
Y_k(\rho)\left(\partial_{\mu}\rho\right)^2
+O(\partial^4)\Bigg\}
\label{derivexp}
\end{split}
\end{equation}
 where     $\vec\phi$     is     a    $N$-component     vector     and
$\rho={\vec\phi}^{\,2}/2$     is    the    $O(N)$     invariant.    In
Eq.(\ref{derivexp}),   $U_k\left(\rho  \right)$  corresponds   to  the
potential part  of $\Gamma_k$  while  $Z_k(\rho)$ and $Y_k(\rho)$
correspond  to   the  field  renormalization   functions.  Thus,  with
$Z_k(\rho)=1$ and $Y_k(\rho)=0$, Eq.(\ref{derivexp}) provides the {\em
ansatz} for  the so-called  Local Potential Approximation  (LPA) where
the    anomalous   dimension   vanishes.    This  kind of  {\em  ansatz}   
has  been successfully used in several cases  among which  the
$O(N)$ \cite{berges02} and Gross-Neveu models
\cite{rosa01,hofling02}.  However, to  deal with  more  complicated models,
{\it e.g.}  with matrix-like  order parameters,  a further  approximation is
almost   unavoidable \cite{tissier00,tissier01,tissier02}.   Indeed,   when   the
symmetry is  lower than $O(N)$,  there are several invariants  and the
number  of  independent  functions  analogous to  $Z_k(\rho,...)$  and
$Y_k(\rho,...)$ grows. In  this case, the integration of  the flow can
be very demanding.  It is then  very convenient to further truncate the
functions $U_k\left(\rho,... \right)$, $Z_k(\rho,...)$ in power series
of $\rho$ and of  all other invariants.

Here, we focus on  the  Ising model,  described by  a  scalar, $\nbZ_2$-invariant
field  theory, considered  as a  toy  model to  study  the  derivative and  field
expansions.   In   this  case,   since  the  only   independent  field
renormalization  function is $Z_k(\rho)$,  the  function $Y_k(\rho)$
can be set to zero. The field truncation then writes:
\be
\left\{
\begin{array}{ll}
U_k(\rho) &=\displaystyle{ \sum_{i=1}^n U_{i,k} (\rho-\rho_0)^i}\\
\\
Z_k(\rho) &=\displaystyle{ \sum_{i=0}^p Z_{i,k} (\rho-\rho_0)^i}
\end{array}
\right.
\label{z0}
\ee
where $\rho_0=\phi_0^2/2$, $\phi_0$ being a particular configuration 
of the field $\phi$. We shall come back on this point later.

The  truncation in fields conveys  two nice properties.  First, with
the  {\em ansatz} (\ref{derivexp}) and  (\ref{z0}), the  RG flow
equation (\ref{eqflot})  leads to  a   finite set  of ordinary  coupled
differential equations  for the coupling constants  $U_{i,k}$'s and $Z_{i,k}$'s
simpler  to solve than  the partial  differential equations  obeyed by
the full functions $U_k(\rho)$   and   $Z_k(\rho)$.  Second,   even   the  lowest   order
approximations,  in   which  only  the  first  nontrivial  terms  of
$U_k(\rho)$ and $Z_k(\rho)$ are  kept, give  a fairly good qualitative
picture of the physics \cite{tetradis94,berges02}.

However, the study of the truncated version of Eq.(\ref{eqflot}) raises  
several important questions:

{\it i}) Does the derivative expansion converge and does it  provide a 
satisfying accuracy at low orders ? The   question  of the convergence 
 of the derivative expansion,  in its
full generality, has not yet been considered and appears to be a major
and open  challenge. In  practice, one is less  interested in
this delicate  question than in the  quality of the  results and their
improvement as the order of  the derivative expansion is increased. In
the case of $O(N)$ models, very accurate results have been obtained at
second  order in  the derivative  expansion.  For  instance, Wetterich
{\it et al.} have shown that handling the full field-dependence of the
potential $U_k(\rho)$ and of the  field renormalization functions $Z_k(\rho)$
and $Y_k(\rho)$ leads to results  that can compete with the world best
estimates,     at     least      for     the     critical     exponent
$\nu$ \cite{berges02}. The  value obtained  for the anomalous
dimension $\eta$  is less
accurate. Its  definition being linked  to the momentum  dependence of
the  two-point  correlation  function,  an accurate  determination  of
$\eta$  probably  requires  higher   order  terms  in  the  derivative
expansion. This question will be investigated in a forthcoming article
 \cite{canet03}.

{\it  ii}) Does  the field  expansion of  $U_k(\rho)$  and $Z_k(\rho)$
converge  and  how  rapidly ?  Once  again, the  general  question  of
convergence has not yet been investigated. Nevertheless, several works
have dealt with field truncations at high order   within  the
LPA \cite{zumbach94b,liao00,litim00,litim01b,aoki98}  or  with  a  field
independent     field-renormalization \cite{tetradis94},    \ie    with
$Z_k(\rho)=Z_{0,k}$. They   suggest  that a  few  orders  suffice to  obtain
reasonably  converged values of  critical exponents.  To our  knowledge, their
computation  using also  an  expansion of  $Z_k(\rho)$  has been  only
studied   in  the   Ising  model   and  using   a   power-law  cutoff
function \cite{aoki98}. In this study,  we  extend this analysis to
two  other  families of cutoff functions, leading to more accurate results.

The questions {\it i}) and {\it ii})  are linked with a corollary issue, which resides 
in the choice of cutoff function. One naturally inquires about its
influence, and in particular: 

{\it iii}) Can the accuracy be improved through the choice of cutoff 
function $R_k$ ? Of course,  when no  truncation is made,  an exact  solution for
$\Gamma[\phi]=\lim_{k\to  0}\Gamma_k[\phi]$, does  not  depend on  the
function  $R_k$ used,  whereas  any kind  of  truncation induces  a
spurious dependence  on it.  One can thus  wonder how to  optimize the
choice of this cutoff function. This question  is not as trivial as it
seems since one  has to decide of  an optimization criterion:  
  rapidity of convergence  of the expansions in powers of  derivatives,   fields   or
amplitudes   \cite{liao00,litim00,litim01,litim01b,litim02} ?   accuracy of
the results ?  sensitivity of  the results with respect to the
cutoff ? We specifically concentrate on these two  latter  issues in the following.

\section{Optimization and Principle  of Minimal Sensitivity}

 Up  to now, attempts  to optimize  nonperturbative RG  equations have
 been mainly worked out in the Polchinski equation \cite{polchinski84}, in particular at second
 order  in  the  derivative  expansion.  For instance,  Ball  {\it  et
 al.} \cite{ball95}  and  Comellas   \cite{comellas98}  have  tried  to
 suppress the  cutoff and  normalization dependence of  the exponents
 $\nu$  and  $\eta$ by  using  the  Principal  of Minimal  Sensitivity
 (PMS) \cite{stevenson81}. We shall not pursue within this framework
 since  it has now been widely recognized that
 the effective average action method is the  most  efficient  way to deal with the
 nonperturbative RG. We will thus consider this latter formalism.

 In the context of the effective average action method, within  the framework 
of LPA, Litim has  proposed to consider  the quantity $C$, defined
by \cite{litim00,litim01, litim01b, litim01c,litim02,litim02b}:
\be
\min_{q^2\ge 0}
\left(\left.\Gamma_k^{(2)}[\phi(q)]\right|_{\phi=\phi_0} + R_k(q)
\right)
= C k^2 
\label{gap}
\ee
where $\Gamma_k^{(2)}[\phi(q)] + R_k(q)$ is the inverse
 of the full regularized propagator  and $C$ parametrizes the gap amplitude.
 According to Litim, the gap is bounded from above and the best cutoff functions are 
those which maximize this gap \cite{litim00,litim01,litim01b,litim01c,litim02,litim02b}:
\be
C_{\hbox{\small opt}}=\max( C) \ \ \ \ {\hbox{when varying $R_k$.}}
\label{critereLitim}
\ee
The idea behind this criterion is that the larger $C$, the more 
stable the truncated RG flow. Indeed, it has been  shown that the maximum
of the gap corresponds to the largest radius of
 convergence of an amplitude expansion. This suggests that the optimal
  selected regulators should have nice properties  such as
    improving the  convergence of the field expansion \cite{litim00,litim01b,litim01c,litim02,litim02b}. Moreover  in \cite{litim01b} it has been shown
 that, within the LPA, the criterion (\ref{critereLitim}) is  also linked to  a
 PMS. 

 At this stage, let us shed the light on some important features
 of  the ``gap criterion''. First Eq.(\ref{critereLitim}) does {\it not} select a unique cutoff
 function: many $R_k$ maximizing  the gap have been exhibited for
 instance in {\cite{litim00}}. Also, the  various  optimized  cutoff
 functions, solutions of Eq. (\ref{critereLitim}), can lead to quantitatively different critical
 exponents  depending on the specific properties of a given cutoff function,
 like its asymptotic  behavior (see below and compare \cite{litim02}
 and \cite{aoki98}). The quality of the results therefore relies on the choice of the type of optimized regulator. 
 Second,  beyond the  LPA,  the implementation of the gap criterion (\ref{critereLitim})
 appears to be  nontrivial. Indeed, the   field renormalization
 function $Z_k(\rho_0)$  induces an implicit $R_k$-dependence in
 $\Gamma_k^{(2)}[\phi(q)]$ that complicates the maximization of the gap. Moreover, it is not completely clear
 whether, beyond the  LPA,  this  criterion would still convey the nice properties
 it shows at the lowest order of the derivative expansion and, in
 particular, its link to a PMS.  As we are specifically concerned here
 with the question of the sensitivity of the results with respect to the
 cutoff function,  we favor
 a method that  directly  probes  the dependence of the critical exponents on the
 cutoff function. We  have decided
 to base our analysis on the PMS, that  can always
 be simply implemented and  has already  proven its efficiency.

 Let us recall how it works. Suppose, for instance,  that we compute a quantity $Q$ 
in an approximate way. The  approximation used  may induce a dependence of $Q$ on a 
parameter -- denoted here  $\alpha$  -- which is spurious. The PMS consists in choosing 
for $\alpha$ the value
$\alpha_{{\tiny PMS}}$ for which $Q$ is stationary:
\be
\left.\frac{d\, Q(\alpha)}{d\, \alpha}\right|_{\alpha_{{\tiny
 PMS}}}=0\ .
\label{pms}
\ee
  One thus expects   that imposing such a constraint,  satisfied by
 $Q$ computed without approximation,  improves the
 approximate determination  of this quantity. The
 obvious drawback of this method is
that Eq.(\ref{pms}) can have many solutions. This worsens if several
quantities are simultaneously studied, and lead to distinct
solutions. An additional criterion is then  necessary to select  a unique one.

We first study  the LPA of the scalar, $\nbZ_2$-invariant 
field theory relevant for the
Ising model. We show that the PMS allows one  to optimize the quality
of the results. We then study the $O(\partial^2)$ approximation of the
derivative expansion and  show that the PMS leads to accurate
results provided we add some new inputs to discriminate the solutions.

\section{The Local Potential Approximation of the Ising model}

Let us recall that the LPA consists in approximating $\Gamma_k$ by:
\be
\Gamma_k[\phi] = \int d^dx\left\{ U_k(\rho) + \frac{1}{2} 
\left(\partial\phi\right)^2\right\}  
\ee
{\it i.e.}  in neglecting the field renormalization.
This {\em ansatz}, once plugged into 
Eq.(\ref{eqflot}), enables to
get the evolution equation for $U_k$. Actually, 
working with dimensionless quantities is necessary to get a fixed point, so that we define:
\be
\left\{
\begin{array}{l}
\displaystyle{r(y)=\frac{R_k(q^2)}{q^2}\ \ \ \ {\hbox{with}}\ \ \ \ 
y=\frac{q^2}{k^2 }}\\
\\
u_k=k^{-d} U_k\\
\\
\tilde\rho=k^{2-d} \rho\ .
\end{array}
\right.
\ee
The RG equation obeyed by $u_k$ writes:
\be
\frac{\partial u_k}{\partial t}= -d u_k +(d-2) \tilde\rho u_k' - v_d L_0^d (w)
\label{evolution_pot}
\ee
where $u_k=u_k(\tilde\rho)$, $v_d^{-1}=2^{d+1} \pi^{d/2} \Gamma(d/2)$, prime means derivation 
with respect to $\tilde\rho$, $w=u_k' +2 \tilde\rho u_k''$, and 
\be
 L_0^d(w)=\int_0^\infty dy\, y^{d/2-1}\frac{2y^2 r'(y)}{y[1+r(y)]+w}.
\ee
The nonperturbative  features of the evolution of the potential are
entirely encoded  in the integral $L_0^d$, called threshold function \cite{tetradis94}. 

 We now study Eq.(\ref{evolution_pot}) within a field truncation:
\be
\begin{array}{ll}
u_k(\tilde\rho) &=\displaystyle{ \sum_{i=1}^n u_i (\tilde\rho-\tilde\rho_0)^i}\ 
\end{array}
\label{developpement-champ}
\ee 
where we have suppressed the index $k$ for the coupling constants.
Once $u_k(\tilde\rho)$ is truncated at a finite order $n$ of the field 
expansion, the field configuration
$\tilde\rho_0$ around which it is expanded matters.
Two configurations have been widely studied: the vanishing
field configuration, $\tilde\rho_0=0$, and the configuration where $u_k(\tilde\rho)$ 
has a  nontrivial  minimum \footnote{Let us notice that although the magnetization vanishes at the 
critical temperature, the dimensionless minimum
of the fixed point potential  $u_*$ does not vanish in general.}:
\be
\left.\frac{\partial u_k}{\partial \tilde\rho}\right|_{\tilde\rho_0}=0.
\ee
All the studies performed  using  field truncations  show that the
convergence properties are improved by expanding around the minimum rather than  around the zero field configuration \cite{morris94c,aoki98}. Therefore, we choose the former.

We also  need to choose  families  of cutoff functions $R_k$ to perform
 calculations. For simplicity, we restrict for now our  study
 to  families  of cutoff functions $R_k$  depending on a single
 parameter.  We extend this to a two-parameter family in section
 \ref{twoparameter}. We consider two usual cutoff functions. The
 first one is the exponential cutoff, which has been often used 
 and constitutes an efficient and robust regulator \cite{tetradis94}.  The other one, the
 theta cutoff, has been introduced by Litim \cite{litim01}. It 
 presents the advantage of leading to  threshold functions that can be
 analytically computed.  We  extend
 theses functions, by  multiplying them by a  factor $\alpha$, to two one-parameter
   families \cite{berges02,litim02}: 
\be
\left\{
\begin{array}{l}
\displaystyle{r_{\exp,\alpha}(y)=\alpha\, \frac{1}{e^y-1}}\\
\\
\displaystyle{r_{\theta,\alpha}(y)=\alpha \left(\frac{1}{y}-1\right)\theta(1-y)}.
\end{array}
\right.
\ee
\begin{figure}[htbp]
\begin{center}
\epsfig{file=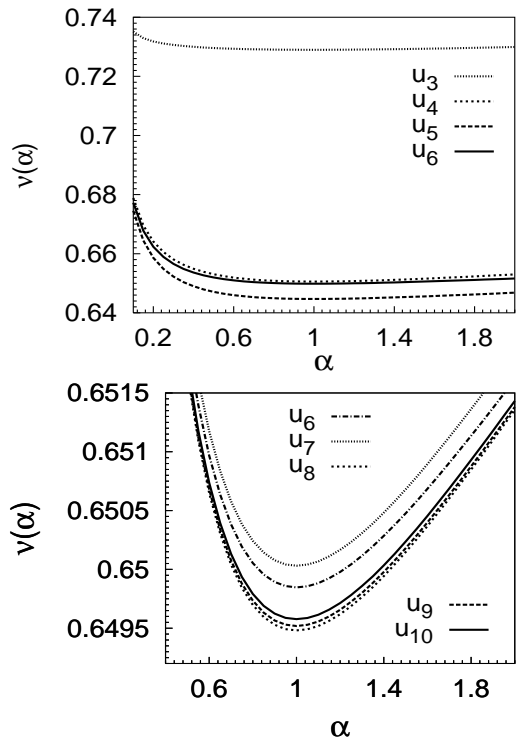,height=10cm,width=7.5cm}
\end{center}
\caption{Curves $\nu(\alpha)$ for the  cutoff $r_{\theta,\alpha}$, for different  truncations of the potential
  $u_k(\tilde\rho)$. Note that, for $n\ge6$ -- lower figure --,  the $\nu$ axis is magnified.}
\label{nuLPA}
\end{figure}

\begin{figure}[htbp]
\begin{center}
\epsfig{file=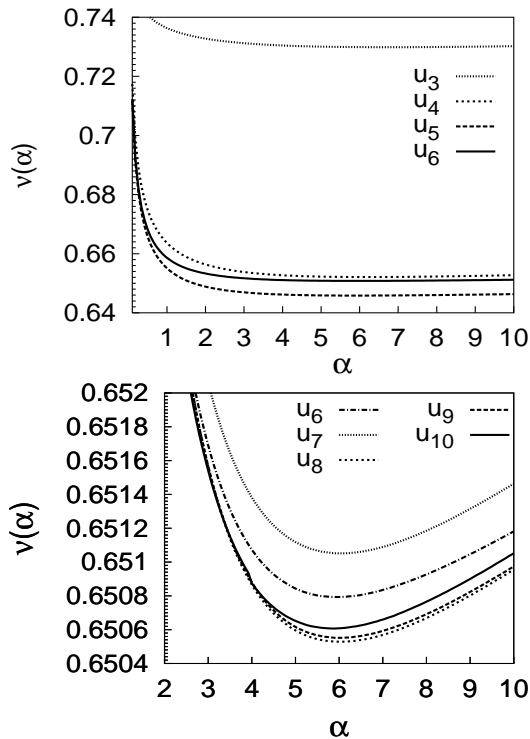,height=10cm,width=7.5cm}
\end{center}
\caption{Curves $\nu(\alpha)$ for the  cutoff $r_{\exp,\alpha}$, for different truncations of the potential
  $u_k(\tilde\rho)$. Note that, for  $n\ge6$ -- lower figure --,  the $\nu$ axis is magnified.}
\label{nuLPAexp}
\end{figure}
Note that both of these cutoff functions can be optimized according to the gap criterion.

For both families, we investigate the $\alpha$ dependence of the
critical exponent $\nu$ over a large range of $\alpha$, for each order
$n$  of the field expansion, up to the tenth power of $\tilde\rho$. We
indeed expect the most  relevant operators to be contained in  the first terms,
and thus the evolution of $\nu$ as a function of the order of the
truncation to be stabilized at, or before,   the tenth order. 

\begin{figure}[htbp]
\begin{center}
\epsfig{file=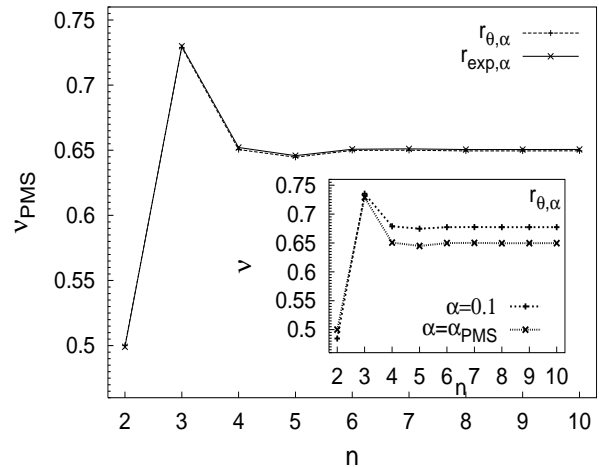,height=8.4cm,width=6.5cm,angle=-90}
\end{center}
\caption{$\nu_{\tiny PMS}$ for $r_{\theta,\alpha}$ and
  $r_{\exp,\alpha}$  as a function of the order $n$ of the field  truncation of $u_k(\tilde\rho)$. The two curves almost superimpose for all $n$. In the insert, $\nu$ is displayed for $r_{\theta,\alpha}$, for both $\alpha_{\tiny PMS}$  and $\alpha=0.1$.}
\label{cvpot}
\end{figure}

We find that, at each order, $\nu(\alpha)$ exhibits a single point of
minimal sensitivity for both cutoff functions. For
$r_{\theta,\alpha}$, (FIG. \ref{nuLPA}), the minimum occurs at
$\alpha_{\tiny PMS}=1$,  as in \cite{litim02},  with an optimized $\nu$ equal to  $\nu(\alpha_{\tiny PMS})=\nu_{\tiny PMS}=0.650$.
For $r_{\exp,\alpha}$, (FIG. \ref{nuLPAexp}), one has  $\alpha_{\tiny
  PMS}=6.03$ and  $\nu_{\tiny PMS}=0.651$ (see Table I). Both cutoff
functions lead to  very similar optimal results for $\nu$, differing
by less than 0.5$\%$ to all orders  $n$, as shown in
FIG. \ref{cvpot}. The converged values of $\nu$ are 
reached  below the percent level in both cases after only a few orders ($n=4$), as expected.

 The insert of FIG. \ref{cvpot}, where the evolution of $\nu$ with
the order $n$ of the field expansion is compared for $\alpha=0.1$ --
chosen  for illustrative purpose -- and $\alpha=\alpha_{\tiny PMS}=1$,
shows that the same convergence level is reached independently of $\alpha$ right from the $n=4$ order, though the asymptotic values of  $\nu(0.1)$ and  $\nu(\alpha_{\tiny PMS})$ differ significantly. This shows that the  rapidity of convergence criterion is helpless here  to select a cutoff. 

 We now compare our results with those obtained through the gap criterion. 
As displayed in FIG. \ref{nuLPA}, $\alpha_{\tiny PMS}=1$ exactly with
 $r_{\theta,\alpha}$  to all orders. For this cutoff function the
 $\alpha_{\tiny PMS}$ value coincides  with that given by the gap
 criterion \cite{litim02}.
For $r_{\exp,\alpha}$, $\alpha_{\tiny PMS}$ converges to 6.03 (see
lower curves in FIG. \ref{nuLPAexp})  whereas the gap  criterion selects an
optimal parameter $\alpha_0=3.92$ \cite{litim00}. In this case, the
two  methods seem to differ. However, since the  variations of $\nu$,
once converged, do not exceed one percent in the whole range $\alpha
\in[\alpha_1\simeq1.2,\alpha_2\simeq74]$, we do not expect the two
methods of optimization to lead to drastically different critical
exponents.        Indeed, $\nu(\alpha_0)=\nu(\alpha_{PMS})$ up to
 $10^{-4}$.   Thus for the  two families of  regulators
 considered  here, the PMS and gap criteria  coincide. It has been
 argued that this  property holds,  within the LPA,  for more general 
 families of regulators \cite{litim01b,litim02}.

 Note that for the exponential cutoff, the standard choice $\alpha = 1$ leads to $\nu=0.658$. This value,  which
 does not correspond to an optimized one, differs by a little bit more than one percent  from
$\nu_{\tiny PMS}$.  For completeness, we also  mention
 that the power-law regulator optimized via the gap   criterion --- 
 $r(y)=y^{-2}$ --- leads to a less accurate  result: $\nu=0.660$
 \cite{aoki98,morris98c}. 

Finally, let us emphasize that the world best value  $\nu=0.6304(13)$
(see Table \ref{tab}) lies below all  curves $\nu(\alpha)$ for both
cutoff functions and  that the PMS solutions for $\nu$ are
minima. Thus, $\nu(\alpha_{PMS})$ is the most accurate value
achievable within each family of cutoff functions studied here.  The
PMS therefore constitutes a powerful method to optimize the cutoff
function in order to reach the best accuracy on the critical exponents.

\begin{table}[htbp]
%\centering
\begin{tabular}{|l|c|c|c|}
\hline
        &Ref.  & $\nu$ & $\eta$ \\
\hline
         &a) & 0.651  &  0  \\
LPA      &b),b')& 0.650 &  0  \\
         &c) & 0.660 &  0  \\
\hline
              &d) & 0.6307 & 0.0467\\
$\partial^2$  &a) & 0.6281 & 0.0443 \\
              &b)  & 0.6260 & 0.0470\\
             &c) & 0.6175 & 0.0542 \\
\hline
7-loop&e) & 0.6304(13) & 0.0335(25)\\
\hline
MC   &f) & 0.6297(5) & 0.0362(8)\\
\hline
Exp.   &g) & 0.636(31) & 0.045(11)\\
      &h) & 0.6298(90)& \\
\hline
\end{tabular}
\caption{Critical exponents of the three dimensional Ising model.
a), b), b'), c) and d) are computed from the effective average action method:
a) with  $r_{\exp,\alpha_{\tiny PMS}}$ (present work);
b) with  $r_{\theta,\alpha_{\tiny PMS}}$ (present work);
b') with $r_{\theta,\alpha=1}$ \cite{litim02};
c)  with power-law cutoff \cite{aoki98,morris98c};
d) with  $r_{\exp,\alpha=1}$ without  field expansion \cite{seide99};
e) from perturbation theory including 7-loop contributions \cite{guida98};
f) from Monte Carlo simulations \cite{hasenbusch01};
g) from experiment in  mixing transition  \cite{muller99};
h) from experiment in liquid-vapor transition (computed from  $3\nu = 2-\alpha$ \cite{haupt99}).}
\label{tab}
\end{table}

\section{Order $\partial^2$ of the derivative expansion}

We now show  how the PMS can be consistently implemented at the order
$\partial^2$ of the derivative expansion for which,
 as far as we know, no optimization procedure has ever been
 implemented within the effective average action method. We dispose of
 two physical quantities candidates for a PMS analysis: $\nu$ and
 $\eta$. We perform both analyses independently, with each cutoff
 function. We show in section \ref{s1} that the PMS allows one to
 improve the accuracy on both exponents. We especially highlight that
 accuracy is not synonymous  of  rapidity of  convergence of the
 field expansion. In section \ref{s2}, we bring out a necessary
 condition for the independent implementation of the two PMS on $\nu$
 and $\eta$ to be consistent. We then check that our results meet this
 condition. In section \ref{s3}, we exhibit cases where, contrary to
 what occurs in the LPA, multiple PMS solutions exist. We show that a
 unique one can be selected thanks to  general arguments. We end up by extending the analysis to a two-parameter family of cutoff functions.  

\subsection{Accuracy of the PMS solution and convergence of the field expansion}
\label{s1}

This section is devoted to showing that the PMS is still, at order $\partial^2$, the appropriate tool to find, within a class of cutoff functions, the one giving the best accuracy. Though it seems counter-intuitive, we emphasize that this cutoff function does not coincide with the one providing the fastest convergence of the field expansion of $Z_k({\rho})$. To this purpose, we implement both PMS independently on $\nu$ and $\eta$, postponing the coherence of this to the next section. 

Working with a nontrivial field renormalization function $Z_k(\rho$), dimensionless {\it and} renormalized quantities are necessary in order to get a fixed point, so that we define:
\be
\left\{
\begin{array}{l}
\displaystyle{r(y)=\frac{R_k(q^2)}{Z_{0,k}\, q^2}\ \ \ \ {\hbox{with}}\ \ \ \ 
y=\frac{q^2}{k^2 }}\\
\\
u_k(\tilde\rho)=k^{-d}\, U_k(\tilde\rho)\\
\\
\tilde\rho=Z_{0,k}\, k^{2-d} \rho.\\
\\
\displaystyle{z_k({\tilde\rho})=\frac{Z_k(\tilde\rho)}{Z_{0,k}}}
\\
\end{array}
\right.
\label{developpementz}
\ee
where $Z_{0,k}$ is defined in Eq.(\ref{z0}).
The RG equation obeyed by  $z_k(\tilde\rho)$ writes \cite{tetradis94,seide99}:
\begin{equation}
\begin{array}{lll}
\displaystyle{\partial z_k\over \partial t}=&\displaystyle
\eta z_k +\tilde\rho\,z_k' (d-2+\eta)+ v_d(z_k'+2\rho
z_k'')L_1^d(w,z_k,\eta)-\\
&\displaystyle 4 v_d \,
\tilde\rho\, z_k'(3 u_k''+2\tilde\rho
u_k''')L_2^d(w,z_k,\eta)-\\
&\displaystyle 2v_d\left(2+{1/ d}\right)\tilde\rho\, (z_k')^2 L_2^{d+2}(w,z_k,\eta)+\\ & \displaystyle
\left({4/d}\right)v_d\tilde\rho\, (3u_k''+2 \tilde\rho u_k''')^2
M_4^d(w,z_k,\eta)+\\
& \displaystyle\left({8/
  d}\right)v_d\,\tilde\rho\, z_k'(3u_k''+2 \tilde\rho
u_k''')M_4^{d+2}(w,z_k,\eta)+\\
& \displaystyle\left({4/ 
  d}\right)v_d\,\tilde\rho\, (z_k')^2 M_4^{d+4}(w,z_k,\eta)
\end{array}
\end{equation}
where $w=u'+2 \tilde\rho u''$, prime means derivative with respect to $\tilde\rho$, and the threshold functions are defined, for $n\ge 1$, by: 

\begin{equation}
\begin{array}{ll}
L_{n}^d(w,z_k,\eta)=\displaystyle n\int_0^\infty dy\; y^{d/2-1}\frac{2y^2r'(y)+\eta y r(y)}{(P(y)+w)^{n+1}}
\end{array}
\end{equation}

\begin{equation}
\begin{array}{ll}
M_{n}^d(w,z_k,\eta)=\displaystyle\int_0^\infty dy\; y^{d/2}\displaystyle\frac{1+r(y)+yr'(y)}
{(P(y)+w)^{n}} \\
\Bigg\{y(1+r(y)+yr'(y))(\eta r(y)+2y r'(y))\bigg(\displaystyle\frac{n}{P(y)+w}\bigg) \\
-2\eta(r(y)+yr'(y))-4y(2r'(y)+yr''(y))\Bigg\}
\end{array}
\end{equation}
where
\begin{align}
P(y)=y(z_k+r(y)).
\end{align}

The anomalous dimension $\eta$ is given by:
\begin{equation}
\eta=-{d\over dt}\ln Z_{0,k}.
\end{equation}

As  previously, we truncate the field renormalization function $z_k(\tilde\rho$) up to the $p$-th power of $\tilde\rho$:
\be
\begin{array}{ll}
z_k(\tilde\rho) &=\displaystyle{ \sum_{i=0}^p z_i (\tilde\rho-\tilde\rho_0)^i}.
\end{array}
\label{developpementz}
\ee
 We use, for the potential $u_k(\tilde\rho)$,   the expansion given in  Eq.(\ref{developpement-champ}), up to the $\tilde\rho^{10}$ term, which represents a very accurate approximation of $u_k(\tilde\rho$) in the vicinity of its minimum as shown in the previous section. We expand $z_k(\tilde\rho)$ up to the ninth power of $\tilde\rho$ which turns out to be sufficient to obtain converged results.
\begin{figure}[htbp]
\begin{center}
\epsfig{file=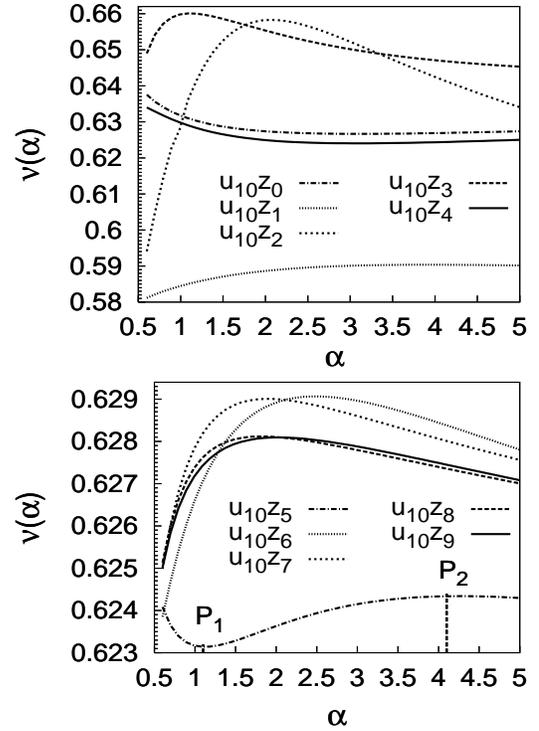,height=10cm,width=7.5cm}
\end{center}
\caption{Curves $\nu(\alpha)$ for  $r_{\exp,\alpha}$,  for different
  truncations  of the field renormalization $z_k(\tilde\rho)$. For
  $p\ge5$ -- lower figure --, the $\nu$ axis is magnified. Note that the
  curve $u_{10}z_5$ shows two extrema [62].}
\label{nuZ}
\end{figure}
\begin{figure}[hbtp]
\begin{center}
\epsfig{file=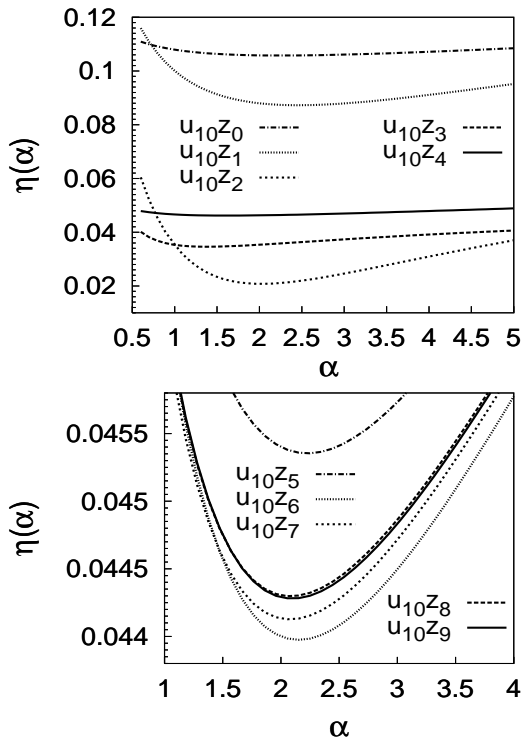,height=10cm,width=7.5cm}
\end{center}
\caption{Curves $\eta(\alpha)$ for  $r_{\exp,\alpha}$, for different
  truncations of the field renormalization $z_k(\tilde\rho)$. For
  $p\ge5$ -- lower figure --, the $\eta$ axis is magnified.}
\label{etaZ}
\end{figure}

 At  each order $p$ of the field expansion of $z_k(\tilde\rho)$, we
 have computed the exponents $\nu$ and $\eta$ as  functions of $\alpha$
 for both cutoff functions  $r_{\theta,\alpha}$ and
 $r_{\exp,\alpha}$. FIG. \ref{nuZ} and FIG. \ref{etaZ} gather the
 curves representing these functions, labelled  $u_{10}z_p$,
 $p=0,\dots,9$, on the example of $r_{\exp,\alpha}$. They are
 displayed on a range of $\alpha$ around the extremum and separated
 in two distinct figures since the $p\ge 5$ curves would be
 superimposed without magnification. This  seems to indicate
 that  the field expansion converges, at least on the whole range of
 $\alpha$ studied. The same conclusion holds for $r_{\theta,\alpha}$,
 with very similar curves (that we therefore do not show), up to the
 important subtlety, discussed in section \ref{s3}, that two PMS
 solutions exist  in this case and that only one has to  be considered. We call $\nu^{\infty}(\alpha)$ and $\eta^{\infty}(\alpha)$ the two limit functions obtained for $p\to\infty$.   In practice, we approximate these functions by those at $p=9$.

 Let us emphasize that for both cutoff functions, {\it i)} the
 rapidity of convergence to $\nu^{\infty}(\alpha)$ and
 $\eta^{\infty}(\alpha)$  and {\it ii)} the asymptotic values
 $\nu^{\infty}$ and $\eta^{\infty}$, depend both on $\alpha$. One can
 thus naturally wonder whether the  values of $\alpha$  for which the
 convergence is the fastest coincide with those for which the
 exponents are the most accurate compared with the world best results.
 We shall  show that this is {\it not} the case contrary to what is
 widely believed.

\subsubsection{Accuracy}

We first bring out  that the PMS exponents are, as in the LPA case,
 the most  accurate ones.
We have determined, for each $p$, the values  $\alpha_{\tiny
 PMS}^{\nu}(p)$ and  $\alpha_{\tiny PMS}^{\eta}(p)$ for which,
 respectively,  $\nu$ and $\eta$ reach their extremum. The
 corresponding exponents are referred to, in the following, as
 $\nu_{\tiny PMS}^p=\nu(\alpha_{\tiny PMS}^{\nu}(p))$ and $\eta_{\tiny
 PMS}^p=\eta(\alpha_{\tiny PMS}^{\eta}(p))$. The obtained PMS
 asymptotic values are $\nu_{\tiny PMS}^\infty = 0.6281$ and
 $\eta_{\tiny PMS}^\infty = 0.0443$ for the exponential cutoff, and
 $\nu_{\tiny PMS}^\infty = 0.6260$ and $\eta_{\tiny PMS}^\infty =
 0.0470$ for the theta cutoff. These values of the exponents are indeed
 the best achievable within each class of cutoff functions studied,
 since the world best value of $\nu$ lies above the sets of curves in
 FIG. \ref{nuZ} and since the extremum is a maximum (and vice versa
 for $\eta$) (see Table \ref{tab} and \footnote{$\nu$ is defined in
 our calculation as minus the inverse of the relevant eigenvalue of
 the stability matrix, linearized around the fixed point. This
 computation differs from the one implemented in \cite{seide99}. It
 generates a  small discrepancy ($0.5\%$) between the (d) value in
 Table \ref{tab} and $\nu(\alpha=1)=0.6272$ computed here, see lower
 curves in FIG. \ref{nuZ}.}). The PMS is thus, as in the LPA case, the
 appropriate tool  to find, among a family of cutoff functions, the one providing the best accuracy.

\subsubsection{Rapidity of convergence}
\begin{figure}[htbp]
\begin{center}
\epsfig{file=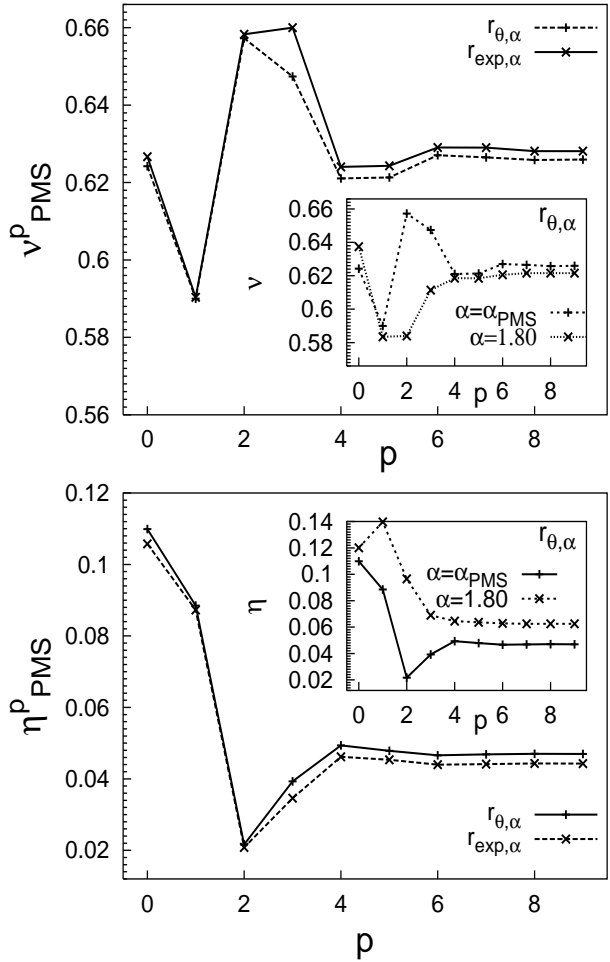,height=13cm,width=8.5cm}
\end{center}
\caption{$\nu_{\tiny PMS}$ and $\eta_{\tiny PMS}$ for both
  $r_{\exp,\alpha}$ and $r_{\theta,\alpha}$ as  functions of the
  order $p$ of the field expansion of $z_k(\tilde\rho)$.  In the
  inserts are displayed, for $r_{\theta,\alpha}$, and  for two
  distinct values of $\alpha$, $\alpha=\alpha_{\tiny PMS}(p)$  and
  $\alpha=1.8$, the critical exponents  $\nu$ (upper insert) and
  $\eta$ (lower insert) as  functions  of $p$.}
\label{cvcin}
\end{figure}

The evolution of  $\nu_{\tiny PMS}^p$ and $\eta_{\tiny PMS}^p$ with
the order $p$ of the field expansion of $z_k(\tilde\rho)$ is
displayed in  FIG. \ref{cvcin}  for both cutoff functions. The
convergence of $\nu_{\tiny PMS}$ and  $\eta_{\tiny PMS}$, at the
percent level, requires at least $p=4$  for both cutoff functions.
However, there exist  values of the parameter $\alpha$, for instance
$\alpha=1.80$ for $r_{\theta,\alpha}$, for which the convergence is
faster than for $\alpha_{\tiny PMS}$.  This is illustrated in the
inserts of  FIG. \ref{cvcin}. Indeed $\eta(\alpha=1.8)$ has already converged at the percent level for $p=3$, but to a different value than $\eta^\infty_{\tiny PMS}$.  Thus, the PMS exponents, which are the most accurate, are not those converging the fastest.  

 We conclude that {\it i)} the PMS leads  to the most accurate exponents within each class of cutoffs studied, {\it ii)} a criterion based on rapidity of convergence of the field expansion would be here misleading since it would select cutoff functions leading to exponents significantly differing from the PMS ones.

\subsection{Consistency condition for independent PMS implementations}
\label{s2}

\begin{figure}[htbp]
\begin{center}
\epsfig{file=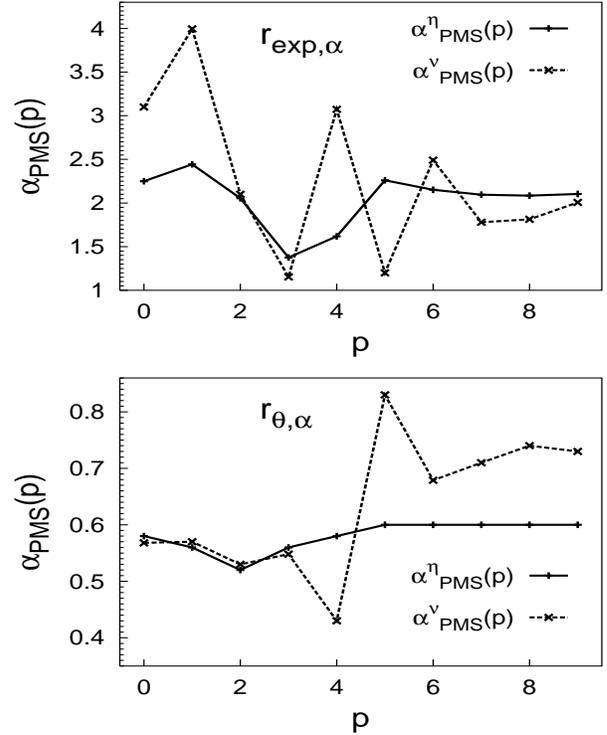,height=10cm,width=8.5cm}
\end{center}
\caption{Plot of $\alpha^\eta_{\tiny PMS}$ and $\alpha^\nu_{\tiny
    PMS}$ as functions of the order $p$ of the field expansion of $z_k(\tilde\rho)$ for both cutoff functions $r_{\exp,\alpha}$ and $r_{\theta,\alpha}$. $\alpha^\eta_{\tiny PMS}$ (resp. $\alpha^\nu_{\tiny PMS}$ ) is the value where lies the PMS extremum of $\eta$, (resp. $\nu$).}
\label{alpha_ms}
\end{figure}
We have implemented and discussed the PMS analyses independently on
$\nu$ and $\eta$ along the previous section. This has naturally led us
to  two distinct PMS values of $\alpha$ at each order $p$, $\alpha_{\tiny
  PMS}^{\nu}(p)$  and $\alpha_{\tiny PMS}^{\eta}(p)$. One can thus
wonder whether it makes sense to compute two different quantities with
two different cutoff functions. We now provide  a natural condition for the whole procedure to be consistent. 
  
Let us notice that since the field expansion seems to converge (as
shown in the previous section), the  two sequences $\alpha_{\tiny
  PMS}^{\nu}(p)$ and $\alpha_{\tiny PMS}^{\eta}(p)$ also
converge. The asymptotic value  $\alpha_{\tiny PMS}^{\eta}(p=\infty)$
(resp. $\alpha_{\tiny PMS}^{\nu}(p=\infty)$) is the one that  achieves
the minimum dependence of the exponent $\eta$ (resp. $\nu$) on the
cutoff function at order $\partial^2$ of the derivative
expansion. There is no reason for them to coincide. However, the
discrepancy between the $\alpha_{PMS}$'s does not matter as long as
choosing one or the other does not change significantly the value of
each exponent. A  consistency condition is thus:
\be
\nu(\alpha_{\tiny PMS}^{\eta}(\infty)) \simeq \nu(\alpha_{\tiny PMS}^{\nu}(\infty))
\label{cond1}
\ee
and
\be
\eta(\alpha_{\tiny PMS}^{\nu}(\infty)) \simeq \eta(\alpha_{\tiny PMS}^{\eta}(\infty))\ .
\label{cond2}
\ee
 Reciprocally, large discrepancies between the values, at the two
 $\alpha_{PMS}$'s, of an exponent would be an indication of a failure
 of convergence. It could be imputed to either a too low order of
 expansion, or to an unappropriate choice of  cutoff functions family.
 
In principle, we should check the consistency over the whole set of
exponents describing the model. Let us however show that once this
condition is satisfied by two independent exponents, it is
automatically by all the others, provided the scaling relations hold
within the chosen truncation scheme (in fields and derivatives).  Let
us first emphasize that it has been observed in all instances where it
has been studied that the scaling relations remain precisely verified
order by order in the field expansion, although the exponents vary
much  with the order. We thus assume that  computing the critical exponents either directly  or from the scaling relations is (almost) equivalent. In this case, an exponent, $\gamma$ for instance, related to $\nu$ and $\eta$ through the scaling relation:
\be
\gamma(\alpha)=\nu(\alpha)\big(2 - \eta(\alpha)\big)
\ee
obviously  verifies for all $\alpha$:
\be
\frac{d\gamma}{d\alpha} = \frac{d\nu}{d\alpha}(2 - \eta) - \nu\frac{d\eta}{d\alpha}.
\label{sca}
\ee
In the simple case where $\alpha_{\tiny PMS}^{\nu}$ coincides with
$\alpha_{\tiny PMS}^{\eta}$, we deduce from Eq.(\ref{sca}) that
$\gamma(\alpha)$ also reaches its extremum for this $\alpha_{\tiny
  PMS}$.  Thus, $\alpha_{\tiny PMS}^{\gamma}=\alpha_{\tiny
  PMS}^{\nu}=\alpha_{\tiny PMS}^{\eta}$ and the consistency is
trivially verified for $\gamma$ also. In the general case where the
$\alpha_{\tiny PMS}$'s are distinct, if they correspond to consistent
exponents $\nu$ and $\eta$ according to Eq.(\ref{cond1}) and
(\ref{cond2}), one is ensured that both exponents are almost
stationary between these two  $\alpha_{\tiny PMS}$'s, provided the
functions  $\nu(\alpha)$ and $\eta(\alpha)$ are smooth enough in this
range. Hence, it follows from Eq.(\ref{sca}) that $d\gamma/d\alpha$
almost vanishes  both at  $\alpha=\alpha_{\tiny PMS}^\eta$ and at
$\alpha=\alpha_{\tiny PMS}^\nu$. This means that $\gamma(\alpha)$  is
also stationary around these points, and thus, $\gamma$ computed from
a PMS analysis should  verify:
\be
\gamma(\alpha_{\tiny PMS}^{\gamma}(\infty)) \simeq \gamma(\alpha_{\tiny PMS}^{\eta}(\infty)) \simeq \gamma(\alpha_{\tiny PMS}^{\nu}(\infty)),
\ee
{\it i.e.} $\gamma$ meets the consistency condition.  Using the same argument for all the other exponents, we deduce that the independent implementations of the PMS on all exponents are consistent once they are for two independent ones.

Let us  now examine our results.
FIG. \ref{alpha_ms} sketches $\alpha_{\tiny PMS}^{\nu}(p)$ and
$\alpha_{\tiny PMS}^{\eta}(p)$  as functions of the order $p$ of the
field truncation, for both cutoff functions $r_{\exp,\alpha}$ and
$r_{\theta,\alpha}$.  Let us set out a few comments. First,
 the  functions $\alpha_{\tiny PMS}^{\nu}(p)$ and  $\alpha_{\tiny
  PMS}^{\eta}(p)$ converge as expected.  On the one hand, $\alpha_{\tiny PMS}^{\eta}(p)$
turns out to be very stable, and roughly converging   as fast as 
$\eta^p_{\tiny PMS}$. This originates in the very peaked shape of the
function $\eta(\alpha)$ (lower curves of FIG. \ref{etaZ}). On the
other hand, $\alpha_{\tiny PMS}^\nu$ shows larger oscillations, 
  due to the flatness of the function
$\nu(\alpha)$ (lower curves of FIG. \ref{nuZ}). It is worth mentioning
that since the exponents have almost  converged at $p=4$ (FIG. \ref{cvcin}), the fluctuations on the corresponding $\alpha_{\tiny PMS}$ values induce negligible variations on them for $p\ge 4$.

 Let us now show  that the independent analyses of $\eta$ and $\nu$
 give consistent results with respect to Eqs. (\ref{cond1}) and (\ref{cond2}). The asymptotic values are approximated by those at  $p=9$. 
The consistency condition is trivially verified for $r_{\exp,\alpha}$  since in this case $\alpha_{\tiny PMS}^{\eta}(\infty) \simeq \alpha_{\tiny PMS}^{\nu}(\infty)$ (see FIG. \ref{alpha_ms}). For $r_{\theta,\alpha}$, we find: 
\be
\begin{array}{l}
|\nu(\alpha_{\tiny PMS}^{\eta}(\infty))- \nu(\alpha_{\tiny PMS}^{\nu}(\infty))| \simeq 10^{-4}\\
\\
|\eta(\alpha_{\tiny PMS}^{\eta}(\infty))- \eta(\alpha_{\tiny PMS}^{\nu}(\infty))| \simeq 6.10^{-4}
\end{array}
\ee
 which are both negligible. Thus, in this case also, the consistency
 condition is fulfilled. We draw the conclusion that the PMS analyses
 have selected a unique  optimal value for each exponent  $\nu$ and $\eta$ although  the corresponding $\alpha_{\tiny PMS}$'s do not coincide. They enable to deduce the remaining critical exponents  as well.

\subsection{Discrimination of multiple PMS extrema}
\label{s3}

\begin{figure}[htbp]
\begin{center}
\epsfig{file=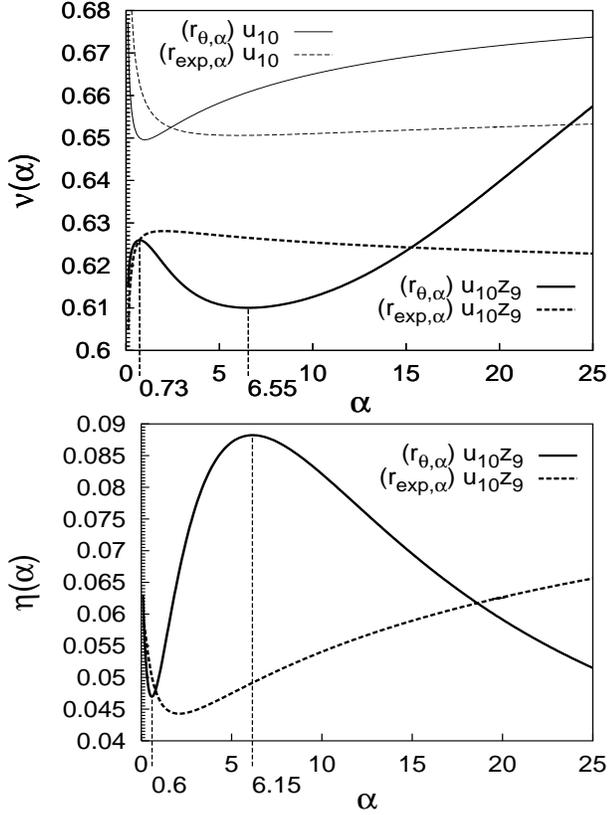,height=11cm,width=8.5cm}
\end{center}
\caption{Curves $\nu(\alpha)$ and $\eta(\alpha)$ for both cutoff
  functions  $r_{\theta,\alpha}$ and  $r_{\exp,\alpha}$ within LPA
  (labelled $u_{10}$) and at $O(\partial^2)$ of the derivative
  expansion (labelled $u_{10}z_9$), for  the maximal truncations of
  $u_k(\tilde \rho)$ and
  $z_k(\tilde \rho)$ computed
  here.  The two PMS extrema for $r_{\theta,\alpha}$ are shown  for both $\nu$ and $\eta$.}
\label{lpaz9}
\end{figure}

The results discussed in section \ref{s1} are associated with a
particular PMS solution while several ones can exist, leading to
significantly different exponents \footnote{Note that several PMS
  solutions can lead to  almost degenerate critical exponents. See for
  instance the curve $u_{10}z_5$  on FIG. \ref{nuZ} for
  $r_{\exp,\alpha}$ where two PMS exist at P$_1$ and P$_2$. In this
  case, whichever point can be selected arbitrarily, since anyway the
  discrepancy between $\nu_{PMS}(P_1)$ and $\nu_{PMS}(P_2)$ is
  negligible (it does not exceed a few tenths of percent here).}.
This  happens for $r_{\theta,\alpha}$, (see FIG. \ref{lpaz9}).  We now
expose the general arguments we used to discriminate between the different PMS solutions.

Suppose that the derivative expansion is studied order by order
without  field truncation (or equivalently that the field expansion is
perfectly converged). If the derivative expansion converges, the
corrections on exponents must be smaller and smaller as the order of
the expansion is increased, at least at sufficiently large order. On
the other hand, as the asymptotic value of any observable is exact, it
must be independent of the  cutoff function. Thus,  for any quantity,
all cutoff functions lead to the same asymptotic  --  exact --  value, although not at the same speed. In practice, the aim is to reach it as fast as possible. This means that, at least beyond a certain order, the best cutoff {\it for the derivative expansion} is the one which leads to the fastest convergence. Note that this is not the case for the field expansion where the rapidity of convergence does not provide a  criterion to discriminate between various PMS solutions.

 Of course, this asymptotic value could be reached only  after large
 fluctuations occuring at first orders, as in the field expansion (see
 FIG. \ref{cvcin}). However, contrary to this case and provided $\eta$
 is not too large, we expect the first orders of the derivative
 expansion to already lead to reliable results. Under this hypothesis,
 we  get two natural  criteria to select a unique PMS solution when
 several exist. The first one  consists in keeping, for each family of cutoff functions,  only the PMS solutions that have a counterpart in the other(s) family(ies), {\it i.e.} that lead to (almost) the same critical exponents. This means in our case that we keep only the PMS solutions that verify (in obvious notations):
\be
\begin{array}{l}
\nu_{\tiny PMS}^{\exp}\simeq \nu_{\tiny PMS}^{\theta}\\
\\
\eta_{\tiny PMS}^{\exp}\simeq \eta_{\tiny PMS}^{\theta}
\end{array}
\ee
 since these exponents are stationary not only inside a family of
 cutoff functions  but also from one family to the other.
The second criterion consists in applying our previous hypothesis of rapid convergence already at order $\partial^2$:
we assume that no large fluctuation occurs between the LPA and $\partial^2$ approximation. We thus select the PMS solution that minimizes, on the exponents, the correction of order $\partial^2$  to the LPA.

Both criteria allow one  to discriminate between the two distinct PMS
 solutions obtained for  $\nu$ and $\eta$ with  $r_{\theta,\alpha}$
 (see the curve $u_{10}z_9$ in  FIG. \ref{lpaz9}).
 They happen to pair for both exponents, at roughly $\alpha_{\tiny PMS}^{\eta} \simeq  \alpha_{\tiny PMS}^{\nu} \simeq 0.7$ and $\alpha_{\tiny PMS}^{\eta} \simeq  \alpha_{\tiny PMS}^{\nu} \simeq 6.5$. According to the second criterion, we exclude the second PMS   solutions located at $\alpha_{\tiny PMS} \simeq 6.5$, which lead for both $\nu$ and $\eta$  to much larger deviations than the first ones compared with the LPA result: $\eta=0$ and $\nu=0.650$ (see FIG. \ref{lpaz9}, curve ($r_{\theta,\alpha}$) $u_{10}$). The first criterion leads to the same choice since {\it i)} we have checked that with $r_{\exp,\alpha}$ only one PMS solution exists for $\nu$ (resp. for $\eta$), and {\it ii)} the corresponding exponent is very similar to the one at the first PMS solution for $\nu$ (resp. for $\eta$) with $r_{\theta,\alpha}$, see Table \ref{tab} and  FIG. \ref{lpaz9}. Thus, our two criteria to select a  unique PMS solution are consistent
\footnote{ Actually, for $\alpha\to\infty$, both $\nu$ and $\eta$ approach an asymptotic value  for $r_{\exp,\alpha}$ that, by extending the notion of  PMS to infinite $\alpha$ could be considered as a second
PMS solution. However, the  values of both $\nu$ and $\eta$ thus obtained -- $\nu(\alpha=\infty)\simeq 0.60$ and  $\eta(\alpha=\infty)\simeq 0.124$ -- are  far from those  at the second PMS of $r_{\theta,\alpha}$ -- $\nu\simeq 0.61$ and  $\eta\simeq 0.088$ --  and therefore cannot be considered as consistent.}.

\subsection{Influence of a second parameter}
\label{twoparameter}

In the previous sections, we have restricted our analyses to the
influence of the parameter $\alpha$, amplitude of the cutoff
functions, on the critical exponents. The optimized results obtained
with the two families of cutoff functions are very close together. It
is thus natural to test the robustness of this result. In  this section
we  investigate the influence of other deformations of the usual
cutoff functions focusing on the exponential cutoff. Two generalizations of
$r_{\exp,\alpha}$ come naturally. They consist in changing  {\it  i)} $\exp y \to \exp \beta y$ and {\it  ii)} $\exp y\to \exp y^\beta$ \cite{liao00,wetterich93,wetterich93c,litim00}. The deformation {\it  i)} reveals actually useless since it is equivalent to a rescaling of the running scale $k$ in $R_k$ which is immaterial. We hence study the two-parameter generalization of $r_{\exp}$:
\be
\displaystyle{r_{\exp,\alpha,\beta}(y)=\alpha \, \frac{1}{e^{y^\beta}-1}}.
\ee

We perform the full PMS analyses of $\nu$ and $\eta$ over the two-parameter space spanned by $\alpha$ and $\beta$,  within the LPA and at order $\partial^2$ of the derivative expansion, for the maximal field truncations of $u_k(\tilde\rho)$
  and $z_k(\tilde\rho)$ considered here. We find a unique two-dimensional PMS for both exponents, and at both orders. It lies at $\alpha_{\tiny PMS}^\eta \simeq  \alpha_{\tiny PMS}^\nu = 2.25$, $\beta_{\tiny PMS}^\nu \simeq \beta_{\tiny PMS}^\eta = 0.98$ and gives $\eta_{\tiny PMS} = 0.04426$ and $\nu_{\tiny PMS} = 0.6281$ at order $\partial^2$. It turns out that our prior choice $\beta=1$ was very close to $\beta_{\tiny PMS}$, and thus the $\alpha$ optimization performed in the previous section already enabled us to almost reach this minimum. The two-parameter PMS exponents thus differ by less than a tenth of percent from those obtained previously (see Table \ref{tab}).

\begin{figure}[htbp]
\begin{center}
\epsfig{file=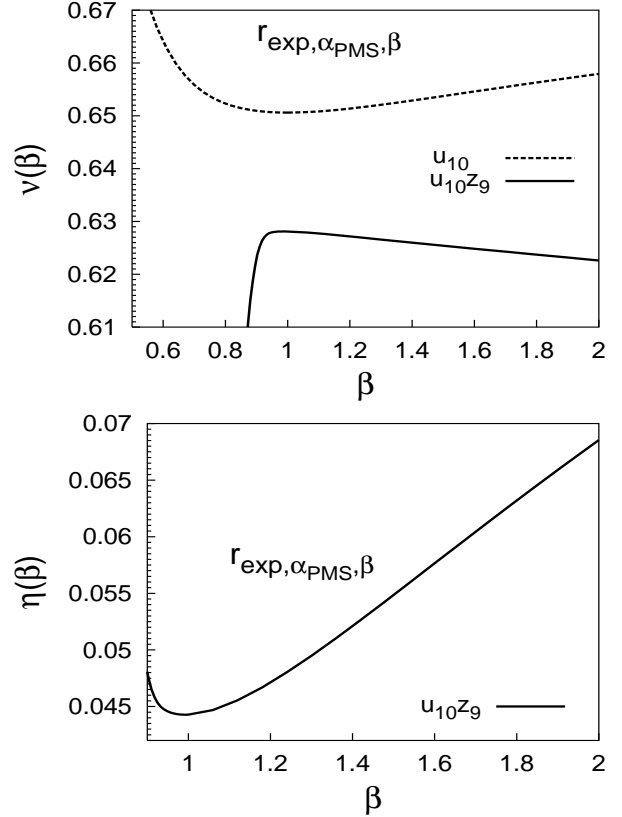,height=11cm,width=8.5cm}
\end{center}
\caption{Curves $\nu(\beta)$ and $\eta(\beta)$ for
  $r_{\exp,\alpha_{\tiny PMS},\beta}$, within LPA (labelled $u_{10}$)
  and at $O(\partial^2)$ of the derivative expansion (labelled
  $u_{10}z_9$) for the maximal truncations of $u_k(\tilde\rho)$
  and $z_k(\tilde\rho)$ considered here.  $\alpha_{\tiny PMS}$ is the value obtained in section \ref{s1}.}
\label{pmsbeta}
\end{figure}

For illustration purpose,  we isolate  in FIG. \ref{pmsbeta} the behavior of the $\beta$ parameter, fixing $\alpha$ to its PMS value determined in section \ref{s1}. It displays the $\eta(\beta)$ and $\nu(\beta)$ functions, for the converged field truncations. Both exponents exhibit a single PMS solution for $\beta$ very close to one ($\beta_{\tiny PMS} = 1.001$ in LPA and $\beta_{\tiny PMS}^\nu = \beta_{\tiny PMS}^\eta = 0.993$ at order $\partial^2$).

As shown in FIG. \ref{pmsbeta}, the $\beta$ dependence of $\nu$ and
$\eta$ is quite sharp. It raises a natural question:  had we
 fixed $\beta$ far from $\beta_{\tiny PMS}$ to perform the $\alpha$
PMS analysis, what would have we obtained ? In other words, would the $\alpha$ optimization have suffice to retrieve exponents close to the two-parameter PMS ones? To investigate this question, we have fixed $\beta=2$, which seems from FIG. \ref{pmsbeta} to alter much $\eta$, and determined $\alpha_{\tiny PMS}^\nu$ and $\alpha_{\tiny PMS}^\eta$. The corresponding exponents are $\eta_{\tiny PMS}^{\beta=2} = 0.05573$, $\nu_{\tiny PMS}^{\beta=2} = 0.6246$ at order $\partial^2$. The discrepancy with the two-parameter PMS exponents is quite significant for $\eta$, whereas the larger exponents -- $\nu$ and the others computed from the scaling relations -- only undergo a few percent variation. 
This originates in the difference of nature of both exponents. On the
one hand, the exponent  $\nu$  is related to the behavior of the mass,
embodied in the minimum of the effective potential. The weakness of
the sensitivity of $\nu$ on the cutoff function, at order
$\partial^2$ of the derivative expansion, suggests that the effective
potential is already well approximated at this order, and thus
provides an accurate determination of $\nu$, close to the exact
value. On the other hand, $\eta$ describes the momentum-dependent part
of the two-spin correlation function, for which the order $\partial^2$
truncation constitutes a very rough {\em ansatz}. Hence, the
determination of $\eta$ is rather poor at this order and improving it
probably requires higher derivative orders. This is directly reflected in the 
non-negligible dependence of $\eta$ on the cutoff function underlined above.

The conclusion to be drawn from this is that, as previously, the PMS
is the appropriate method to select, among a class of cutoff functions,
the one that achieves  the best accuracy, in so far as it minimizes
the distance to the world best values for both exponents and at both
orders. Moreover, the PMS reveals itself  all the more crucial that the variations with respect to a given parameter are large.

\section{Conclusion}

We have implemented the Principle of Minimal Sensitivity to improve
critical exponents within the framework of the nonperturbative RG. We
have shown that it always allows to reach the most accurate results
achievable in the class of cutoff functions under scrutiny.  Within
the  LPA, the PMS exponents turn out to almost coincide with those
obtained through the principle of  maximization of the gap, 
and  the method is  easily generalizable at order $\partial^2$. 

Two main drawbacks are usually attributed to the implementation of the
PMS: {\it i)}  several solutions of the PMS can exist and render its
implementation ambiguous, {\it ii)} it is not clear whether it indeed
improves the results. We have shown on the example of the Ising model,
that within the context of the effective average action method, these
drawbacks either can be circumvented or do not exist at all. We have
indeed brought out that a unique solution of the PMS can always be
selected, thanks to  very reasonable criteria, and furthermore this
solution represents the most accurate determination of the critical
exponents. The PMS thus  appears as a safe and powerful method to optimize the results obtained in the nonperturbative RG context. An important and rather unexpected aspect of our analysis is that the rapidity of convergence of the field expansion is not optimal where the accuracy is.

Let us also  emphasize that, even within a rather modest truncation involving the potential expansion up to order $\tilde\rho^5$  and the field renormalization expansion up to order $\tilde\rho^4$, the accuracy reached on $\nu$ is below the percent level compared with the world best results. This suggests that, with the same kind of computational complexity,  a comparable accuracy can be achieved for more complicated models. 

Finally, the determination of $\eta$ is  poorer, which is to be imputed to the roughness of the {\it ansatz} to  describe the full momentum dependence of the two-spin correlation function. Improving it is likely to require inclusion of terms of order $\partial^4$. This will be investigated in \cite{canet03}.   

\acknowledgments

We are  particularly indebted to  D. Litim for very interesting and
fruitful  discussions as well as  for having communicated to us  data prior to publication. The LPTHE and the GPS are  Unit\'{e}s Mixtes du CNRS, respectively UMR 7589 and UMR 7588.

\end{document}